\documentclass[conference]{IEEEtran}

\usepackage{amsmath}
\usepackage{amssymb}
\usepackage[dvips]{graphicx}
\usepackage{setspace}
\usepackage{epsfig}
\usepackage{cite}
\usepackage{color}
\usepackage{algorithm,algpseudocode,caption}
\usepackage{xcolor}
\usepackage{lipsum}

\newtheorem{theo}{Theorem}
\newtheorem{prop}{Proposition}
\newcommand{\figsize}{0.45}
\newcommand{\figsizee}{0.9}

\newenvironment{thisnote}{\par\color{black}}{\par}

\begin{document}
\IEEEoverridecommandlockouts
\title{Effective Capacity in Multiple Access Channels with Arbitrary Inputs}
\author{Marwan Hammouda, Sami Akin, and J\"{u}rgen Peissig\\Institute of Communications Technology\\Leibniz Universit\"{a}t Hannover\\Email: \{marwan.hammouda, sami.akin, and peissig\}@ikt.uni-hannover.de\thanks{This work was partially supported by the European Research Council under Starting Grant--306644.}}
\date{}

\maketitle

\begin{abstract}
In this paper, we consider a two-user multiple access fading channel under quality-of-service (QoS) constraints. We initially formulate the transmission rates for both transmitters, where the transmitters have arbitrarily distributed input signals. We assume that the receiver performs successive decoding with a certain order. Then, we establish the effective capacity region that provides the maximum allowable sustainable arrival rate region at the transmitters' buffers under QoS guarantees. Assuming limited transmission power budgets at the transmitters, we attain the power allocation policies that maximize the effective capacity region. As for the decoding order at the receiver, we characterize the optimal decoding order regions in the plane of channel fading parameters for given power allocation policies. In order to accomplish the aforementioned objectives, we make use of the relationship between the minimum mean square error and the first derivative of the mutual information with respect to the power allocation policies. Through numerical results, we display the impact of input signal distributions on the effective capacity region performance of this two-user multiple access fading channel.
\end{abstract}

\section{Introduction}
With the growth in wireless networks, recent years witnessed a large body of research on cooperative transmissions \cite{tao2012overview}. The researchers in some of these studies concentrated on multiple access transmission scenarios and investigated these scenarios from an information-theoretic perspective \cite{biglieri2007multiple,tse1998multiaccess,gupta2006power,knopp1995information,vishwanath2001optimum,viswanath2001asymptotically}. For instance, the authors in \cite{tse1998multiaccess} defined the ergodic capacity region for multiple access fading channels and derived the optimal resource allocation policies that maximize this region. Similarly, addressing the optimal power allocation policies that achieve any point on the capacity region boundary subject to a sum-power constraint, Gupta \emph{et al.} studied Gaussian parallel (non-interacting) multiple access channels \cite{gupta2006power}. Moreover, taking the vector fading multiple access channels, the authors examined the dynamic resource allocation policies as an important means to increase the sum capacity in uplink synchronous code-division multiple-access systems \cite{viswanath2001asymptotically}.

It is very well known that the use of discrete and finite constellation diagrams is required for input signaling in many practical systems. Different than the above studies where the authors consider Gaussian input signaling, the authors in \cite{harshan2011two} researched two-user Gaussian multiple access channels with finite input constellations. Equivalently, the authors in \cite{lozano2006optimum} considered parallel Gaussian channels with arbitrary inputs as well. They investigated the optimal power allocation that maximizes the mutual information subject to an average power constraint by exploiting the relationship between the mutual information and the minimum mean-square error (MMSE), which was established in \cite{guo2005mutual}. Furthermore, the authors in \cite{nguyen2010outage} explored the optimal power policies that minimize the outage probability over block-fading channels with arbitrary input distributions that were subject to both peak and average power constraints. Power allocation policies for a two-way relay channel with arbitrary inputs were studied in low and high signal-to-noise ratio regimes. In another line of research, the author studied the multiple access multiple-input multiple-output channels, and showed the relationship between the input-output mutual information and the MMSE \cite{ghanem2012mac}.

In the meantime, since the current wireless systems require data transmission with strict constraints on delay performance, cross-layer design concerns have become of interest to many system designers. Therefore, quality-of-service (QoS) requirements regarding buffer overflow and delay have been addressed in wireless communications studies regarding the Data-Link and Physical layers. In that regard, effective capacity was established as a measure to indicate the maximum sustainable rate at a transmitter queue by a given service (channel) process \cite{wu_negi}. Consequently, effective capacity has been investigated in several different transmission scenarios \cite{tang,gursoy,ak_gur_4}. More recently, Ozcan \emph{et al.} studied the effective capacity of point-to-point channels and derived the optimal power allocation policies to maximize the system throughput by employing arbitrary input distributions under average power constraints.

In this paper, we focus on a two-user multiple access transmission scenario in which transmitters apply arbitrarily distributed input signaling under average power constraints and QoS requirements that are imposed as buffer overflow and delay probabilities. Our analysis can be easily expanded to multiple access scenarios with more than two transmitters. Our main contributions can be sorted as follows: Defining the effective capacity region by employing the effective capacity of each transmitter, we provide the optimal power allocation policies under an average transmission power constraint. We make use of the relationship between the mutual information and the MMSE in obtaining the power allocation policies. Furthermore, we attain the optimal decoding order that is administered at the receiver regarding the interplay between the channel fading coefficients.

\section{System Description}\label{sec:system_description}
\begin{figure*}
\centering 
\includegraphics[width=\figsizee\textwidth]{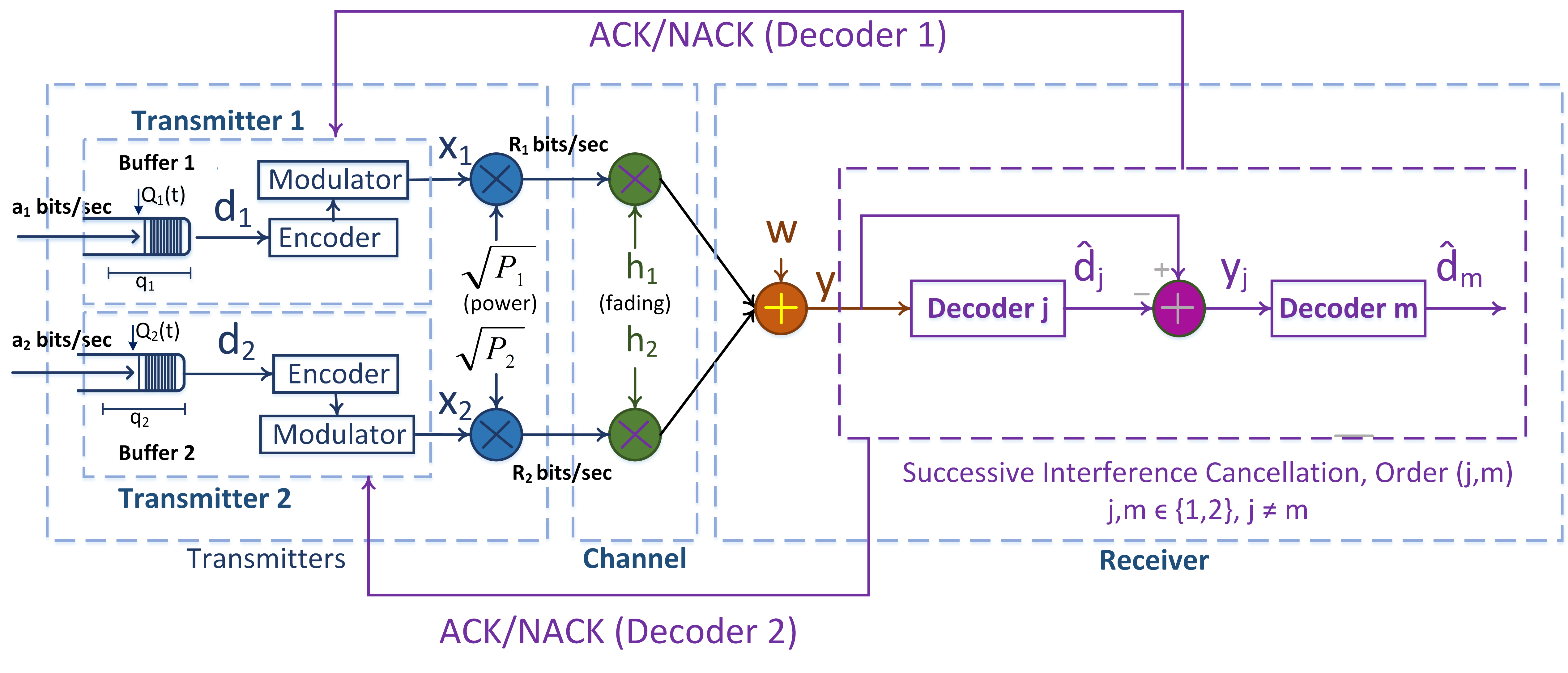}
\caption{Channel model. We consider a two-user multiple access channel in which two transmitters are communicating with a single receiver. Each transmitter has a data buffer, and the receiver performs successive interference cancellation with a certain order.}\label{System_Model}
\end{figure*}

\subsection{Channel Model}
We consider a multiple access channel scenario in which two transmitters send data to one common receiver as seen in Figure \ref{System_Model}. We initially assume that the data arrive at both transmitters from a source (or sources), and they are stored in the transmitters' data buffers before being conveyed into the wireless channel. Then, each transmitter divides the available data into data packets and performs the encoding, modulation and transmission of each packet in frames of $T$ seconds. If a packet is received and decoded correctly by the receiver, the receiver sends a positive acknowledgment (ACK) to the corresponding transmitter (i.e., the transmitter that sends the packet), and the transmitter removes the packet from its buffer. Otherwise, the receiver sends a negative ACK (NACK) to the corresponding transmitter, and the transmitter resends the same packet. Thus, we impose certain QoS requirements in each transmitter buffer in order to control the buffer violation probabilities.

During the transmission in the channel, the input-output relation at time instant $t$ is given as 
\begin{align*}
\label{input_output_general}
y(t) = \sqrt{P_1(t)} h_1(t) x_1(t) + \sqrt{P_2(t)} & h_2(t) x_2(t)+ w(t),
\end{align*}
for $t = 1,2,\cdots$. Above, $x_1(t)$ and $x_2(t)$ are the channel inputs at the corresponding transmitters (i.e., Transmitter 1 and 2, respectively, in Fig. 1), and $y(t)$ is the channel output at the receiver. $P_1(t)$ and $P_2(t)$ are the instantaneous power allocation policies employed by Transmitter 1 and 2, respectively, with the following average power constraint:
\begin{equation}\label{avg_constraint_combined}
\mathbb{E}\{P_1(t)\}+\mathbb{E}\{P_2(t)\}\leq\overline{P},
\end{equation}
where $\overline{P}$ is finite. Moreover, $w(t)$ denotes the zero-mean, circularly symmetric, complex Gaussian random variable with a unit variance, i.e., ${E}\{|w|^2\}=1$. The noise samples $\{w(t)\}$ are independent and identically distributed. Meanwhile, $h_1(t)$ and $h_2(t)$ represent the fading coefficients between Transmitter 1 and the receiver, and Transmitter 2 and the receiver, respectively. The magnitude squares of the fading coefficients are denoted by $z_1(t) = |h_1(t)|^2$ and $z_2(t) = |h_2(t)|^2$ with finite averages, i.e., ${E}\{z_{1}\}<\infty$ and ${E}\{z_{2}\}<\infty$. We consider a block-fading channel, and assume that the fading coefficients stay constant for a frame duration of $T$ seconds and change independently from one frame to another. The channel coefficients, $h_1$ and $h_2$, are perfectly known to the receiver and both transmitters, and hence, each transmitter can adapt its transmission power policy accordingly. We finally note that the available transmission bandwidth is $B$ Hz. In the rest of the paper, we omit the time index $t$ unless otherwise needed for clarity.

\subsection{Achievable Rates}
We can express the instantaneous achievable rate between the transmitters and the receiver by invoking the mutual information between the inputs at the transmitters, i.e., $x_1$, $x_2$, and the output at the receiver, i.e., $y$. Hence, given that the instantaneous channel fading values, $h_{1}$ and $h_{2}$, are available at the transmitters and the receiver, the instantaneous achievable rate can be given as \cite{gallager1968information}
\begin{equation}
\label{mutual_total}
\mathcal{I}(x_1,x_2;y) = \mathbb{E} \left \{\log_{2}\frac{f_{y|x_1,x_2}(y|x_1,x_2)}{ f_y(y)} \right \},
\end{equation}
where $f_y(y) = \sum_{x_1,x_2} p(x_1,x_2) f_{y|x_1,x_2}(y|x_1,x_2)$ is the marginal probability density function (pdf) of the received signal $y$ and
\begin{equation*}
f_{y|x_1,x_2}(y|x_1,x_2) = \frac{1}{\pi} e^{-|y-\sqrt{\alpha_1 \overline{P}} h_1 x_1- \sqrt{\alpha_2 \overline{P}} h_2 x_2|^2 }.
\end{equation*}
Above, we consider the normalized power allocation policies: $\alpha_1 = \frac{P_1}{\overline{P}}$ and $\alpha_2 = \frac{P_2}{\overline{P}}$.

We assume that the receiver performs successive interference cancellation with a certain order $(j,m)$ for $j,m \in \{1,2\}$ and $j \neq m$. The decoding order depends on the channel conditions, i.e., the magnitude squares of channel fading coefficients, $z_{1}$ and $z_{2}$. In particular, the receiver initially decodes $x_j$ while treating $x_m$ as noise, and then subtracts $x_{j}$ from the received signal $y$ and decodes $x_m$. Let $\mathcal{Z}$ be the region of the $(z_1,z_2)$-space where the decoding order is (2,1). Then, $\mathcal{Z}^{c}$, which is the complement of $\mathcal{Z}$, is the region where the decoding order is (1,2). Now, we can express the instantaneous transmission rates for each transmitter as follows:
\begin{align}
&r_1(z_1,z_2)=
\begin{cases}
\mathcal{I}(x_1;y_1),& \mathcal{Z},\\
\mathcal{I}(x_1;y), &\mathcal{Z}^{c},
\end{cases}\label{R_1}
\intertext{and}
&r_2(z_1,z_2)=
\begin{cases}
\mathcal{I}(x_2;y), &\mathcal{Z},\\
\mathcal{I}(x_2;y_2),&\mathcal{Z}^{c},
\end{cases}\label{R_2}
\end{align}
where 
\begin{equation}\label{y_1_2}
\begin{aligned}
& y_1 = \sqrt{\alpha_1 \overline{P}} h_1 x_1 + w, \\
& y_2 = \sqrt{\alpha_2 \overline{P}} h_2 x_2 + w.
\end{aligned}
\end{equation}
The decoding regions can be determined in such a way to maximize the objective throughput. Furthermore, we have
\begin{align*}
\mathcal{I}(x_j;y_j)=\mathbb{E} \left \{\log_{2}\frac{f_{y_j|x_j}(y_j|x_j)}{ f_{y_{j}}(y_{j})}\right\},
\end{align*}
where $f_{y_j}(y_j) = \sum_{x_j} p(x_j) f_{y_j|x_j}(y_j|x_j)$ is the marginal pdf of $y_{j}$ and
\begin{equation*}
f_{y_j|x_j}(y_j|x_j)=\frac{1}{\pi} e^{-|y_j-\sqrt{\alpha_j \overline{P}} h_j x_j|^2 }.
\end{equation*}
 
\subsection{Effective Capacity}
Recall that the data packets are stored in the buffers of the transmitters until they are reliably decoded by the receiver. Thus, the delay and buffer overflow concerns are of interest for system designers. Therefore, we concentrate on the data arrival processes, i.e., $a_{1}$ and $a_{2}$ in Fig. \ref{System_Model}, and we propose the effective capacity that provides us the maximum constant arrival rate that a given service (channel) process can support in order to guarantee a desired statistical QoS specified with the QoS exponent $\theta$ \cite{wu_negi}.

Now, let $Q$ be the stationary queue length, then we can define the decay rate of the tail distribution of the queue length $Q$ as 
\begin{equation*}
\theta = - \lim_{q \to \infty} \frac{\log \text{Pr}(Q\geq q)}{q}.
\end{equation*}
Therefore, for large $q_{max}$ we can approximate the buffer violation probability as $\text{Pr}(Q \geq q_{max}) \approx e^{- \theta q_{max}}$. Based on this relation, we can see that large $\theta$ indicates stricter QoS constraints, while smaller $\theta$ implies looser constraints. For a discrete-time, stationary and ergodic stochastic service process $r(t)$, the effective capacity is given by
\begin{equation*}
-\lim_{t \to \infty} \frac{1}{\theta t} \log_e \mathbb{E} \{e^{-\theta S(t)}\},
\end{equation*}
where $S(t) = \sum_{\tau=1}^t r(\tau)$. Hence, the effective capacity identifies the asymptotic decay rate of buffer occupancy, and it can be considered as the dual of the effective bandwidth \cite{chang1994effective}.

In the aforementioned multiple access transmission scenario, each transmitter has its own buffer to store the data, and it has its own QoS requirements. Therefore, we denote the decay rate of Transmitter 1 and Transmitter 2 by $\theta_1$ and $\theta_2$, respectively. Noting that the transmission bandwidth is $B$ Hz, the block duration is $T$ seconds, and the channel fading coefficients change independently from one transmission frame to another, we can express the effective capacity of each transmitter, i.e., the maximum sustainable data arrival rate at Transmitter $j$, in bits/sec/Hz as
\begin{equation}
\label{effective_capacity_j}
-\frac{1}{\theta_j T B} \log_e \mathbb{E} \left \{ e^ {-\theta_j T B r_j(z_1,z_2)}\right\} \quad j\in\{1,2\},
\end{equation}
where the expectation is taken over the $(z_1,z_2)$-space. Now, invoking the definition given in \cite{qiao2013achievable}, we express the effective capacity region of the given multiple access transmission scenario as follows:
\begin{align}\label{effective_rate_region}
\mathcal{C}_E(\Theta)=&\bigcup_{r_1,r_2}\Big\{{C(\Theta)}\geq {\bf 0}:\nonumber\\
&{C_j(\theta_{j})\leq -\frac{1}{\theta_j T B} \log_e \mathbb{E} \left \{ e^ {-\theta_j T B r_j(z_1,z_2)}\right\}}\Big\},
\end{align}
where $\Theta = [\theta_1,\theta_2]$, and ${C(\Theta)}=[C_{1}(\theta_{1}),C_{2}(\theta_2)]$ is the vector of the effective capacity values.
 
\section{Performance Analysis}
In this section, we focus on maximizing the effective capacity region defined in (\ref{effective_rate_region}) under the QoS guarantees required at each transmitter and the average total power constraint defined in (\ref{avg_constraint_combined}). Noting that the effective capacity region is convex \cite{qiao2012transmission}, our objective turns out to be maximizing the boundary surface of the region, which can be characterized by the following optimization problem \cite{tse1998multiaccess}:
\begin{equation}\label{obtimization_objective}
\max_{\substack{\mathcal{Z},\mathcal{Z}^{c}\\\mathbb{E}\{P_{1}\}+\mathbb{E}\{P_{2}\}\leq\overline{P}}} \lambda_1 {C}_1(\theta_1) + \lambda_2 {C}_2(\theta_2),
\end{equation}
for $\lambda_1,\lambda_2\in[0,1]$ such that $\lambda_1+\lambda_2=1$. In order to solve this optimization problem, we first obtain the power allocation policies in defined decoding regions $\mathcal{Z}$ and $\mathcal{Z}^{c}$, and then we provide the optimal decoding regions.

\subsection{Optimal Power Allocation}
Here, we study the optimal power allocation policies that solve the optimization problem in (\ref{obtimization_objective}) in given decoding regions $\mathcal{Z}$ and $\mathcal{Z}^{c}$. In the subsequent result, we provide the following proposition that gives us the optimal power allocation policies:
\begin{prop}\label{proposition_1}
The optimal normalized power allocation policies, $\alpha_1$ and $\alpha_2$, that solve the optimization problem in (\ref{obtimization_objective}) are the solutions of the following equalities:
\begin{align}
&\frac{\lambda_1}{\psi_1}e^{-\theta_1 TBr_1(z)} \frac{dr_1(z)}{d \alpha_1} + \frac{\lambda_2}{\psi_2} e^{-\theta_2 TBr_2(z)} \frac{dr_2(z)}{d \alpha_1}  = \varepsilon, \label{optimal_alpha_1_Z}\\
&\frac{\lambda_2}{\psi_2} e^{-\theta_2 TBr_2(z)} \frac{dr_2(z)}{d \alpha_2}  = \varepsilon, \label{optimal_alpha_2_Z}
\end{align}
for $z = (z_1,z_2) \in \mathcal{Z}$, and
\begin{align}
&\frac{\lambda_1}{\psi_1} e^{-\theta_1 TBr_1(z)} \frac{dr_1(z)}{d \alpha_1} = \varepsilon, \label{optimal_alpha_1_Z_c}\\
&\frac{\lambda_1}{\psi_1} e^{-\theta_1 TBr_1(z)} \frac{dr_1(z)}{d \alpha_2} + \frac{\lambda_2}{\psi_2} e^{-\theta_2 TBr_2(z)} \frac{dr_2(z)}{d \alpha_2} = \varepsilon, \label{optimal_alpha_2_Z_c}
\end{align}
for $z \in \mathcal{Z}^c$. Above, $\psi_1 = \mathbb{E}_z \big \{ e^{-\theta_1 TBr_1(z)} \big \}$, $\psi_2 = \mathbb{E}_z \big \{ e^{-\theta_2 TBr_2(z)} \big \}$, and  $\varepsilon$ is the Lagrange multiplier of the average power constraint in (\ref{avg_constraint_combined}).
\end{prop}
\emph{Proof:} See Appendix \ref{app:proposition_1}. $\hfill{\square}$

Above, the derivatives of the transmission rates with respect to the corresponding normalized power allocation policies are given as
\begin{align*}\label{R_derivative}
\frac{dr_1(z)}{d \alpha_1}=\begin{cases}
\frac{d\mathcal{I}(x_1;y_1)}{d\alpha_1},& \mathcal{Z},\\
\frac{d\mathcal{I}(x_1;y)}{d\alpha_1}, &\mathcal{Z}^{c},
\end{cases}\\
\frac{dr_2(z)}{d \alpha_2}=\begin{cases}
\frac{d\mathcal{I}(x_2;y)}{d\alpha_2}, &\mathcal{Z},\\
\frac{d\mathcal{I}(x_2;y_2)}{d\alpha_2},&\mathcal{Z}^{c},
\end{cases}
\end{align*}
and 
\begin{equation*}
\frac{dr_m(z)}{d \alpha_j} = \frac{d \mathcal{I}(x_j;y)}{d \alpha_j} - \frac{d \mathcal{I}(x_j;y_j)}{d \alpha_j}
\end{equation*}
for $m,j\in\{1,2\}$ and $m\neq j$.

In the following theorem, we provide the derivatives of the mutual information expressions with respect to the normalized power allocation policies:
\begin{theo}\label{theo:der_alpha}
Let, $h_1$, $h_2$, and $\overline{P}$ be given. In the multiple access transmission scenario described in Section \ref{sec:system_description}, the first derivative of the mutual information between $x_{j}$ and $y$ with respect to the power allocation policy, $\alpha_j$, is given by
\begingroup
\allowdisplaybreaks
\begin{align}
&\frac{d\mathcal{I}(x_j;y)}{d \alpha_j} =  \overline{P} z_j \text{MMSE}(x_j;y) \notag \\
&\hspace{0.9cm} + \overline{P} \sqrt{\frac{\alpha_m}{\alpha_j}} \text{Re}\left(h_j h_m^{*}  \mathbb{E}\left\{x_{j}x_{m}^{*}-\hat{x}_j(y) \hat{x}_m^{*}(y)\right\}\right),\label{d_I_tot_alpha}
\end{align} 
and similarly, the derivative of the mutual information between $x_{j}$ and $y_{j}$ with respect to $\alpha_j$ is given by
\begin{equation}
\frac{d\mathcal{I}(x_j;y_j)}{d \alpha_j} = \overline{P} z_j \text{MMSE}(x_j;y_j),\label{d_I_tot_alpha_2}
\end{equation}
\endgroup
for $j,m \in \{1,2\}$, $j \neq m$, and $(\cdot)^{*}$ is the complex conjugate operation. In (\ref{d_I_tot_alpha}), the MMSE expression is given as
\begin{align*}
\text{MMSE}(x_j;y) =  1 - \frac{1}{\pi} \int \frac{\big |\sum_{x_j} x_j  p(x_j) f_{y|x_j}(y|x_j) \big |^2}{f_{y}(y)} \mathrm{d}y,
\end{align*}
and the MMSE estimates of the channel inputs are
\begin{equation*}
\hat{x}_j(y) = \frac{\sum_{x} x_j  p(x) f_{y|x}(y|x)}{f_{y}(y)}.
\end{equation*}
Similarly, the MMSE expression in (\ref{d_I_tot_alpha_2}) is obtained by
\begin{align*}
\text{MMSE}(x_j;y_j) =  1 - \frac{1}{\pi} \int \frac{\big |\sum_{x_j} x_j  p(x_j) f_{y_j|x_j}(y_j|x_j) \big |^2}{f_{y_j}(y_j)}\mathrm{d}y_j,
\end{align*}
where $y_{1}$ and $y_{2}$ are as given in (\ref{y_1_2}).
\end{theo}
\emph{Proof:} See Appendix \ref{app:der_alpha}. $\hfill{\square}$

As seen in (\ref{optimal_alpha_1_Z})-(\ref{optimal_alpha_2_Z_c}), closed-form solutions for $\alpha_{1}$ and $\alpha_{2}$ cannot be obtained easily which is mainly due to the cross-relation between $\alpha_{1}$ and $\alpha_{2}$. For instance, $\alpha_1$ is a function of $\alpha_{2}$ as observed in (\ref{optimal_alpha_1_Z}) for $z \in \mathcal{Z}$, whereas $\alpha_{2}$ is a function of $\alpha_{1}$ as seen in (\ref{optimal_alpha_2_Z_c}) for $z \in \mathcal{Z}^c$. Therefore, we need to employ numerical techniques which consist of iterative solutions.

In the following, we wrap up the above steps into an iterative solution with two algorithms that can be used to obtain the optimal power policies in given decoding regions. In Algorithm \ref{algorithm_1}, we obtain the optimal normalized power allocation policies $\alpha_1$ and $\alpha_2$.

\begingroup\captionof{algorithm}{}\label{algorithm_1}
	\begin{algorithmic}[1] 
    \State Given $\lambda_1$, $\lambda_2$, $\mathcal{Z}$ and $\mathcal{Z}^c$;
    \State Initialize $\psi_1$, $\psi_2$;\label{int_psi}
    \While {True} 
    \State Initialize $\varepsilon$;
    \State Initialize $\alpha_1$; \label{kappa}
    \While {True}
    \If {$z \in \mathcal{Z}$}
    \State For given $\alpha_1$, compute the optimal $\alpha_2$ by solving (\ref{optimal_alpha_2_Z}) \label{alg:alpha_2_Z};
    \State For computed $\alpha_2$, compute the optimal $\alpha_1^{\star}$ by solving (\ref{optimal_alpha_1_Z}) \label{alg:alpha_1_Z};
    \Else 
    \State For given $\alpha_1$, compute the optimal $\alpha_2$ by solving (\ref{optimal_alpha_2_Z_c}) \label{alg:alpha_2_Z_c};
    \State For computed $\alpha_2$, compute the optimal $\alpha_1^{\star}$ by solving (\ref{optimal_alpha_1_Z_c}) \label{alg:alpha_1_Z_c};
    \EndIf
    \If {$|\alpha_1 - \alpha_1^{\star}| \leq \epsilon$ for small $\epsilon > 0$}
    \State break;
    \Else
    \State Set $\alpha_1 = \alpha_1^{\star}$;
    \EndIf
    \EndWhile
    \State Check if the average power constraint in (\ref{avg_constraint_combined}) is satisfied with quality;
    \State If not, update $\varepsilon$ and return to Step \ref{kappa}
    \State Compute $\psi_1^{\star} = \mathbb{E}_z \big \{ e^{-\theta_1 n r_1(z)} \big \}$ and  $\psi_2^{\star} = \mathbb{E}_z \big \{ e^{-\theta_2 n r_2(z)} \big \}$
    \If {$|\psi_1 - \psi_1^{\star}| \leq \epsilon$ and $|\psi_2 - \psi_2^{\star}| \leq \epsilon$}
    \State {\bf break};
    \Else
    \State Set $\psi_1 = \psi_1^{\star}$ and $\psi_2 = \psi_2^{\star}$;
    \EndIf
    \EndWhile
	\end{algorithmic}
\endgroup

Given $\lambda_j$ and $\psi_j$ for $j \in \{1,2\}$, it is shown in \cite{ozcan2014qos} that both (\ref{optimal_alpha_2_Z}) and (\ref{optimal_alpha_1_Z_c}) has at most one solution. We can further show that (\ref{optimal_alpha_1_Z}) has at most one solution for $\alpha_{1}$ when $\alpha_{2}$ is given, and that (\ref{optimal_alpha_2_Z_c}) has at most one solution for $\alpha_{2}$ when $\alpha_{1}$ is given. Then, we can guarantee that Steps \ref{alg:alpha_2_Z}, \ref{alg:alpha_1_Z}, \ref{alg:alpha_2_Z_c} and \ref{alg:alpha_1_Z_c} in Algorithm \ref{algorithm_1} will converge to a single unique solution. It is also clear that (\ref{optimal_alpha_1_Z}) and (\ref{optimal_alpha_1_Z_c}) are monotonically decreasing functions of $\alpha_1$, and (\ref{optimal_alpha_2_Z}) and (\ref{optimal_alpha_2_Z_c}) are monotonically decreasing functions of $\alpha_2$. Hence, in region $\mathcal{Z}$, we first obtain $\alpha_{2}$ by solving (\ref{optimal_alpha_2_Z}), and then we find $\alpha_{1}$ by solving (\ref{optimal_alpha_1_Z}) after inserting $\alpha_{2}$ into (\ref{optimal_alpha_1_Z}). Similarly, in region $\mathcal{Z}^c$, we first obtain $\alpha_{1}$ by solving (\ref{optimal_alpha_1_Z_c}), and then we find $\alpha_{2}$ by solving (\ref{optimal_alpha_2_Z_c}) after inserting $\alpha_{1}$ into (\ref{optimal_alpha_2_Z_c}). We can employ bisection search methods to obtain $\alpha_{1}$ and $\alpha_{2}$. In the above approach, when either $\alpha_{1}$ or $\alpha_{2}$ becomes negative, we set it to zero.

\subsection{Optimal Decoding Order}
Following the optimal power allocation policies, we identify the optimal decoding order regions. We initially note that when there are no QoS requirements, i.e., $\theta_1=\theta_2=0$, the effective capacity region is reduced to be the ergodic capacity region. The authors in \cite{jindal2004duality} showed that the ergodic capacity region is maximized when the symbol of the transmitter with the strongest channel is decoded first. Principally, when $z_{j}>z_{m}$, the symbol of Transmitter $j$ is decoded first, and then the symbol of Transmitter $m$ is decoded. Furthermore, the authors in \cite{qiao2013achievable} considered a special case and set $\theta_1 = \theta_2 = \theta$ for $\theta > 0$. Then, they derived the optimal decoding order that maximizes the effective capacity region. However, their result is based on the assumption of Gaussian input signaling. Nevertheless, obtaining the optimal decoding order regions is a difficult task when $\theta_1\neq\theta_2$ and arbitrary input distribution is employed. In the following, we provide the optimal decoding order regions given that the transmitters have the equal queue decay rates, i.e., $\theta_1=\theta_2$, and they employ arbitrary input distributions.

\begin{theo}\label{theo:optimal_order}
Let $h_1$, $h_2$, and $\overline{P}$ be given. Define $z_2^\star$ for any given $z_1\geq0$, such that the decoding order is (2,1) when $z_2>z_2^{\star}$, and it is (1,2) otherwise for the given $z_1$. In the multiple access transmission scenario described in Section \ref{sec:system_description}, with arbitrary input distributions and the normalized power allocation policies at the transmitters, the optimal $z_2^\star$ for any given $z_1$ value is the solution of the following equality:
\begin{equation*}\label{optimal_oeder}
\mathcal{I}(x;y|z_1,z_2^\star) = \mathcal{I}(x_1;y_1|z_1) + \mathcal{I}(x_2;y_2|z_{2}^\star).
\end{equation*}
\end{theo} 
\emph{Proof:} See Appendix \ref{app:optimal_order}. $\hfill{\square}$	

\begin{figure}
\centering 
\includegraphics[width=\figsize\textwidth]{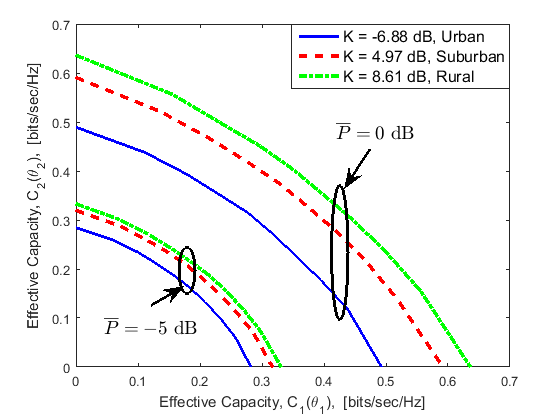}
\caption{Effective capacity region, $C_1(\theta_1)$ vs. $C_2(\theta_2)$, when BPSK input signaling is employed for different values of $\overline{P}$ and $K$.}\label{fig_1}
\end{figure}
\section{Numerical Results}
 
In this section, we present the numerical results. Throughout the paper, we set the available channel bandwidth to $B = 100$ Hz and the transmission duration block to $T = 1$ sec.. We further assume that $h_1$ and $h_2$ are independent of each other and set $\mathbb{E}\{|h_1|^2\} = \mathbb{E}\{|h_2|^2\} = 1$. Unless indicated otherwise, we set the QoS exponents $\theta_1 = \theta_2 = 0.01$. We define the signal-to-noise ratio with $\frac{\overline{P}}{{E}\{|w|^2\}}=\overline{P}$ where ${E}\{|w|^2\}=1$.  

We initially consider binary phase shift keying (BPSK) at both transmitters, and we plot the effective capacity region in Fig. \ref{fig_1}. We have the results for different values of the signal-to-noise ratio, $\overline{P}$, and $K$. Recall that when $K=0$, the channel fading has a Rayleigh distribution, i.e., there is not a strong line-of-sight propagation path between the transmitters and the receiver. On the other hand, when $K>0$, there is a line-of-sight path between the transmitters and the receiver, and the line-of-sight propagation path becomes dominant with increasing $K$\footnote{$K$ is the ratio of the power in the line-of-sight component to the total power in the non-line-of-sight components in a channel. Therefore, the ratio of the power in the line-of-sight component to the total channel power is defined as $\nu = \frac{K}{K + 1}$. It is shown in \cite{jeong2012mimo} that the  empirical means of $K$ are -6.88 dB, 8.61 dB and 4.97 dB for urban, rural and suburban environments, respectively, at 781 MHz.}. As expected, with increasing $K$, the effective capacity region broadens. Moreover, we observe the broadening of the effective capacity region with increasing $\overline{P}$ more clearly.

\begin{figure}
\centering 
\includegraphics[width=\figsize\textwidth]{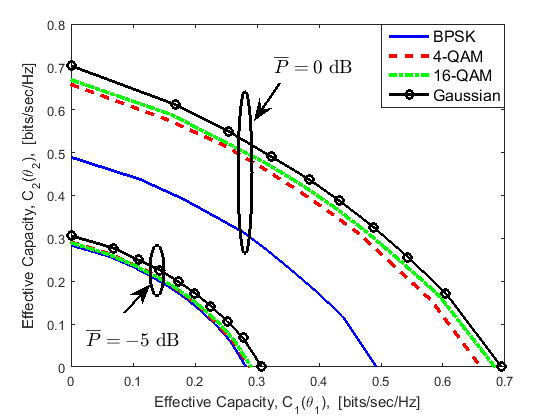}
\caption{Effective capacity region, $C_1(\theta_1)$ vs. $C_2(\theta_2)$, considering different input signaling for $K = -6.88$ dB and different values of $\overline{P}$.}\label{fig_2}
\end{figure}

\begin{figure}
\centering 
\includegraphics[width=\figsize\textwidth]{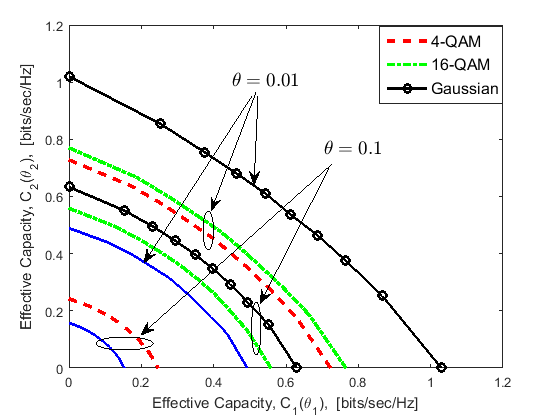}
\caption{Effective capacity region, $C_1(\theta_1)$ vs. $C_2(\theta_2)$, considering different input signaling for $K = -6.88$ dB, $\overline{P} = 5$ dB and different values of $\theta = \theta_1 = \theta_2$.}\label{fig_3}
\end{figure}

Setting $K = -6.88$ dB, we plot the effective capacity region for different $\overline{P}$ values and signal modulation methods such as BPSK, quadrature amplitude modulation (QAM) and Gaussian distributed signaling in Fig. \ref{fig_2}. We can easily notice that Gaussian input signaling has the best performance for both $\overline{P}=-5	$ dB and $\overline{P}=0$ dB, while BPSK has the lowest performance. However, the performance gap is reduced with decreasing $\overline{P}$. Furthermore, we investigate the effect of the QoS exponent, $\theta$, on the effective capacity region in Fig. \ref{fig_3}. Here, we set $\overline{P} = 5$ dB and $K = -6.88$ dB, and compare the effective capacity region for different modulation techniques. As clearly seen, increasing $\theta$ results in a decrease in the effective capacity region since the system is subject to stricter QoS constraints. We can further observe that the performance gaps among the modulation techniques are smaller with increasing $\theta$. We finally display the effective capacity region for transmitters having different modulation methods than each other in Fig. \ref{fig_4}. We can clearly notice that the transmitter with an input signal of higher modulation order can sustain higher effective capacity.

\section{Conclusion}\label{conclusion}
In this paper, we have investigated the optimal power allocation policies that maximize the effective capacity region of a two-user multiple access channel with arbitrarily distributed input signals. We have formulated the relationship between the MMSE and the first derivative of the mutual information with respect to the power allocation policies. We have provided an algorithm that determines the optimal normalized power allocation policies. We have established the optimal decision region boundaries for successive interference cancellation at the receiver for given power allocation policies. Through numerical techniques, we have shown that the line-of-sight propagation path can significantly improve the effective capacity performance. We have further justified that the Gaussian input signaling has better performance and that the performance gap increases in higher signal-to-noise ratio regime. 

\appendix
\subsection{Proof of Proposition \ref{proposition_1}}\label{app:proposition_1}
Let us rewrite (\ref{effective_capacity_j}) for Transmitter 1 as
\begingroup
\allowdisplaybreaks
\begin{align}
\label{C_1}
C_1(\theta_1)  = &\frac{-1}{\theta_1 TB} \log_e \bigg \{ \mathbb{E}_{\mathcal{Z}} \{e^{- \theta_1 TB\mathcal{I}(x_1;y_1)}\} \nonumber\\&+ \mathbb{E}_{\mathcal{Z}^c} \{e^{- \theta_1 TB \mathcal{I}(x_1;y)}\} \bigg \} \notag \\
= &\frac{-1}{\theta_1 TB} \log_e \psi_1,
\end{align}
\endgroup
and for Transmitter 2 as
\begin{align}
\label{C_2}
C_2(\theta_2)  =& \frac{-1}{\theta_2 TB} \log_e \bigg \{ \mathbb{E}_{\mathcal{Z}} \{e^{- \theta_2 TB \mathcal{I}(x_2;y)}\} \nonumber\\&+ \mathbb{E}_{\mathcal{Z}^c} \{e^{- \theta_2 TB \mathcal{I}(x_2;y_2)}\} \bigg \} \notag \\
 = &\frac{-1}{\theta_2 TB} \log_e \psi_2.
\end{align}
Since the objective function in (\ref{effective_rate_region}) is convex and the constraint (\ref{avg_constraint_combined}) is linear with respect to $\alpha_1$ and $\alpha_2$, we can use the Lagrangian method to solve the optimization problem (\ref{obtimization_objective}). We can form the Lagrangian as 
\begin{equation*}
\label{Lagrangian}
\begin{aligned}
& \mathcal{B} =  \lambda_1 C_1(\theta_1)  + \lambda_2 C_2(\theta_2) \\ 
& \hspace{1.0cm} - \varepsilon \{\mathbb{E}_{z \in \mathcal{Z}}\{\alpha_1 + \alpha_2\} + \mathbb{E}_{z \in \mathcal{Z}^c}\{\alpha_1 + \alpha_2\} -1 \},
\end{aligned}
\end{equation*} 
where $\varepsilon$ is the Lagrangian multiplier. Now, taking the derivatives of $\mathcal{B}$ with respect to $\alpha_1$ and $\alpha_2$ and setting them to zero, we obtain (\ref{optimal_alpha_1_Z}) and (\ref{optimal_alpha_2_Z_c}), respectively, when $z \in \mathcal{Z}$, and (\ref{optimal_alpha_1_Z_c}) and (\ref{optimal_alpha_2_Z}), respectively, when $z \in \mathcal{Z}^c$.
\begin{figure}
\centering 
\includegraphics[width=\figsize\textwidth]{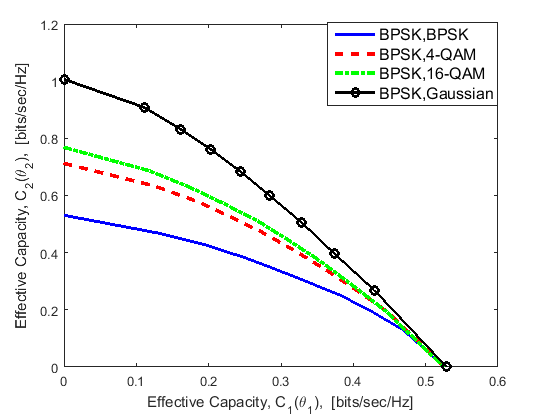}
\caption{Effective capacity region, $C_1(\theta_1)$ vs. $C_2(\theta_2)$, considering mixed input signaling for $K = -6.88$ dB, $\overline{P} = 0$ dB and $\theta_1 = \theta_2 = 0.01$.}\label{fig_4}
\end{figure}
\begin{thisnote} 
\subsection{Proof of Theorem \ref{theo:der_alpha}}\label{app:der_alpha}
Recall that $\alpha_{1}=\frac{P_{1}}{\overline{P}}$ and $\alpha_{2}=\frac{P_{2}}{\overline{P}}$, and
\begin{equation}\label{pdf_y}
f(y) = \sum_x p(x) f(y|x),
\end{equation}
where $x = (x_1,x_2)$. Since our analysis is performed in the complex plane, we can express $f(y|x)$ as
\begin{align}
\label{marginal_pdf}
f(y|x) = \frac{1}{\pi} \text{exp} & \bigg\{-\bigg(y_r - \sqrt{P_1} c_{1r} - \sqrt{P_2} c_{2r}\bigg)^2 \notag \\
& -\bigg(y_i - \sqrt{P_1} c_{1i} - \sqrt{P_2} c_{2i}\bigg)^2 \bigg \},
\end{align}
where $y = y_r + jy_i$, $h_1 x_1 = c_{1r}+jc_{1i}$ and $h_2 x_2 = c_{2r}+jc_{2i}$. The derivative of the pdf with respect to $P_1$ is given as 
\begin{equation*}
\frac{d f(y|x)}{d P_1} = \frac{f(y|x)}{\sqrt{P_1}}  \begin{pmatrix} c_{1r} & c_{1i} \end{pmatrix} \begin{pmatrix} y_r - \sqrt{P_1} c_{1r} - \sqrt{P_2} c_{2r}\\y_i - \sqrt{P_1} c_{1i} - \sqrt{P_2} c_{2i} \end{pmatrix},
\end{equation*}
and 
\begin{equation}\label{der_marginal}
\frac{d f(y|x)}{d y} = \dot{f}(y|x) = -2f(y|x)\begin{pmatrix} y_r - \sqrt{P_1} c_{1r} - \sqrt{P_2} c_{2r}\\y_i - \sqrt{P_1} c_{1i} - \sqrt{P_2} c_{2i} \end{pmatrix}.
\end{equation}
Hence, we have 
\begin{equation}
\label{d_p_y}
\frac{d f(y|x)}{d P_1} = \frac{-1}{2 \sqrt{P_1}} \begin{pmatrix} c_{1r} & c_{1i} \end{pmatrix} \dot{f}(y|x).
\end{equation}
Now, we can express \footnote{This is based on the known relation $\mathcal{I}(x;y) = \mathcal{I}(x_j;y) + \mathcal{I}(x_m;y_m)$ for $j,m \in \{1,2\}$ and $j \neq m$ \cite{tse2005fundamentals}.}
\begin{equation}
\label{step_1}
\frac{d\mathcal{I}(x_1;y)}{d P_1} = \frac{d\mathcal{I}(x;y)}{d P_1} - \underbrace{\frac{d \mathcal{I}(x_2;y_2)}{d P_1} }_{=0}.
\end{equation}
Invoking the marginal pdf $f(y|x)$ in (\ref{marginal_pdf}), we can rewrite the mutual information $\mathcal{I}(x;y)$ expressed in (\ref{mutual_total}) as 
\begin{equation*}
\mathcal{I}(x;y) = - \log(\pi e) - \int f(y) \log(f(y)) \mathrm{d}y. 
\end{equation*} 
Consequently, we can write (\ref{step_1}) as 
\begin{align}
& \frac{d \mathcal{I}(x_1;y)}{d P_1} = \frac{d \mathcal{I}(x;y)}{d P_1} =  - \frac{d}{d P_1} \int f(y) \log(f(y)) \mathrm{d}y \notag, \\
& \hspace{1.7cm}  = - \int [1 + \log(f(y))] \frac{d f(y)}{d P_1} \mathrm{d}y. \label{step_2}
\end{align}
Substituting (\ref{pdf_y}) and (\ref{d_p_y}) in (\ref{step_2}), we obtain 
\begin{align}
\label{step_3}
& \frac{d \mathcal{I}(x_1;y)}{d P_1} = \notag \\ 
& \frac{1}{2 \sqrt{P_1}} \sum_x p(x) \begin{pmatrix} c_{1r} & c_{1i} \end{pmatrix} \int [1 + \log(f(y))] \dot{f}(y|x) \mathrm{d}y.
\end{align}
Let $m = [1 + \log(f(y))]$ and $d n = \dot{f}(y|x) \mathrm{d}y$, then the integration in (\ref{step_3}) can be evaluated using integration by part such that $\int m d n = mn \int n dm$. By noting that $m n = 0$ as $y \to \infty$, we can write (\ref{step_3}) as
\begin{equation}
\label{Step_3_a}
\frac{d \mathcal{I}(x_1;y)}{d P_1} = \frac{-1}{2 \sqrt{P_1}} \sum_x p(x) \begin{pmatrix} c_{1r} & c_{1i} \end{pmatrix} \int \frac{f(y|x)}{f(y)} \dot{f}(y|x) \mathrm{d}y.
\end{equation}
By plugging (\ref{der_marginal}) in (\ref{Step_3_a}), we have

\begingroup
\allowdisplaybreaks
\begin{align*}
&\frac{d \mathcal{I}(x_1;y)}{d P_1}\\
& = \frac{1}{\sqrt{P_{1}}}\sum_{x}p(x)\begin{pmatrix}a_{1r}&a_{1i}\end{pmatrix}\int\frac{f(y|x)}{f(y)}\sum_{x}p(x)f(y|x)\\
& \hspace{1.1cm} \times\begin{pmatrix}y_{r}-\sqrt{P_{1}}a_{1r}-\sqrt{P_{2}}a_{2r}\\y_{i}-\sqrt{P_{1}}a_{1i}-\sqrt{P_{2}}a_{2i}\end{pmatrix}dy,\\
& = \frac{1}{\sqrt{P_{1}}}\sum_{x}p(x)\begin{pmatrix}a_{1r}&a_{1i}\end{pmatrix}\int\frac{f(y|x)}{f(y)}\sum_{x}p(x)f(y|x)\begin{pmatrix}y_{r}\\y_{i}\end{pmatrix}dy\\
& \hspace{0.3cm} -\sum_{x}p(x)\begin{pmatrix}a_{1r}&a_{1i}\end{pmatrix}\int\frac{f(y|x)}{f(y)}\sum_{x}p(x)f(y|x)\begin{pmatrix}a_{1r}\\a_{1i}\end{pmatrix}dy\\
& \hspace{0.3cm} -\frac{\sqrt{P_{2}}}{\sqrt{P_{1}}}\sum_{x}p(x)\begin{pmatrix}a_{1r}&a_{1i}\end{pmatrix}\nonumber\\&\hspace{3.10cm}\times\int\frac{f(y|x)}{f(y)}\sum_{x}p(x)f(y|x)\begin{pmatrix}a_{2r}\\a_{2i}\end{pmatrix}dy,\\
&=\frac{1}{\sqrt{P_{1}}}\sum_{x}p(x)\begin{pmatrix}a_{1r}&a_{1i}\end{pmatrix}\int f(y|x)\begin{pmatrix}y_{r}\\y_{i}\end{pmatrix}dy\\
& \hspace{0.3cm} -\int f(y)\begin{pmatrix}\widehat{a}_{1r}&\widehat{a}_{1i}\end{pmatrix}\begin{pmatrix}\widehat{a}_{1r}\\\widehat{a}_{1i}\end{pmatrix}dy\\
& \hspace{0.3cm} -\frac{\sqrt{P_{2}}}{\sqrt{P_{1}}}\int f(y)\begin{pmatrix}\widehat{a}_{1r}&\widehat{a}_{1i}\end{pmatrix}\begin{pmatrix}\widehat{a}_{2r}\\\widehat{a}_{2i}\end{pmatrix}dy,\\
&=\frac{1}{\sqrt{P_{1}}}\sum_{x}p(x)\begin{pmatrix}a_{1r}&a_{1i}\end{pmatrix}\begin{pmatrix}\sqrt{P_{1}}a_{1r}+\sqrt{P_{2}}a_{2r}\\\sqrt{P_{1}}a_{1i}+\sqrt{P_{2}}a_{2i}\end{pmatrix}\\
& \hspace{0.3cm} -\mathbb{E}\left\{\widehat{a}_{1r}^{2}+\widehat{a}_{1i}^{2}\right\}-\frac{\sqrt{P_{2}}}{\sqrt{P_{1}}}\mathbb{E}\left\{\widehat{a}_{1r}\widehat{a}_{2r}+\widehat{a}_{1i}\widehat{a}_{2i}\right\},\\
&=\mathbb{E}\left\{a_{1r}^{2}+a_{1i}^{2}\right\}+\frac{\sqrt{P_{2}}}{\sqrt{P_{1}}}\mathbb{E}\left\{a_{1r}a_{2r}+a_{1i}a_{2i}\right\}\\
& \hspace{0.3cm} -\mathbb{E}\left\{\widehat{a}_{1r}^{2}+\widehat{a}_{1i}^{2}\right\}-\frac{\sqrt{P_{2}}}{\sqrt{P_{1}}}\mathbb{E}\left\{\widehat{a}_{1r}\widehat{a}_{2r}+\widehat{a}_{1i}\widehat{a}_{2i}\right\},\\
&=|h_{1}|^{2}\mathbb{E}\left\{|x_{1}|^{2}\right\}+\frac{\sqrt{P_{2}}}{\sqrt{P_{1}}}\text{Re}\left(h_{1}h_{2}^{\star}\mathbb{E}\left\{x_{1}x_{2}^{\star}\right\}\right)\\
& \hspace{0.3cm} -|h_{1}|^{2}\mathbb{E}\left\{|\widehat{x}_{1}|^{2}\right\}-\frac{\sqrt{P_{2}}}{\sqrt{P_{1}}}\text{Re}\left(h_{1}h_{2}^{\star}\mathbb{E}\left\{\widehat{x}_{1}\widehat{x}_{2}^{\star}\right\}\right),\\
&=|h_{1}|^{2}\text{MMSE}(x_{1};y)+\frac{\sqrt{P_{2}}}{\sqrt{P_{1}}}\text{Re}\left(h_{1}h_{2}^{\star}\mathbb{E}\left\{x_{1}x_{2}^{\star}-\widehat{x}_{1}\widehat{x}_{2}^{\star}\right\}\right).
\end{align*}
\endgroup
Now, by applying the chain rule $\frac{d (\cdot)}{d \alpha_1} = \overline{P} \frac{d (\cdot)}{d P_1}$, we have
\begingroup
\allowdisplaybreaks
\begin{align*}
& \frac{d \mathcal{I}(x_{1};y)}{d\alpha_{1}}= \overline{P} z_1 \text{MMSE}(x_{1};y)\nonumber\\
& \hspace{2.3cm} + \overline{P} \sqrt{\frac{\alpha_{2}}{\alpha_1}}\text{Re}\left(h_{1}h_{2}^{\star}\mathbb{E}\left\{x_{1}x_{2}^{\star}-\widehat{x}_{1}\widehat{x}_{2}^{\star}\right\}\right).
\end{align*}
\endgroup
Note that when the data of Transmitter 2 is subtracted from the received signal $y$, i.e, $\alpha_2 = 0$, the above equation is reduced to
\begingroup
\allowdisplaybreaks
\begin{align*}
\frac{d \mathcal{I}(x_{1};y_1)}{d\alpha_{1}}= \overline{P} z_1 \text{MMSE}(x_{1};y_1).
\end{align*}
\endgroup

In a similar way, we can show that 
\begingroup
\allowdisplaybreaks
\begin{align*}
& \frac{d \mathcal{I}(x_{2};y)}{dP_{2}}= \overline{P} z_2 \text{MMSE}(x_{2};y)\nonumber\\
& \hspace{2.3cm}+ \overline{P} \sqrt{\frac{\alpha_{1}}{\alpha_2}} \text{Re}\left(h_{2}h_{1}^{\star}\mathbb{E}\left\{x_{2}x_{1}^{\star}-\widehat{x}_{2}\widehat{x}_{1}^{\star}\right\}\right).
\end{align*}
\endgroup
and 
\begingroup
\allowdisplaybreaks
\begin{align*}
\frac{d \mathcal{I}(x_{2};y_2)}{d\alpha_{2}}= \overline{P} z_2 \text{MMSE}(x_{2};y_2).
\end{align*}
\endgroup

\subsection{Proof of Theorem \ref{theo:optimal_order}}\label{app:optimal_order}
We initially start by expressing the effective capacity of each user in  (\ref{C_1}) and (\ref{C_2}) in the integration form and with respect to $z_2^{\star}$ and $\theta_1 = \theta_2 = \theta$ as 

\begin{align*}
& C_1(\theta,z_2^{\star}) \notag \\
& = \frac{-1}{\theta n} \log_e \bigg ( \int_0^{\infty} \int_{z_2^{\star}}^{\infty} e^{- \theta n \mathcal{I}(x_1;y_1|z_1)} p_{{z}}(z) \mathrm{d}z_2 \mathrm{d}z_1\notag \\
& + \int_0^{\infty} \int_{0}^{z_2^{\star}} e^{- \theta n \mathcal{I}(x;y|z_1,g(z_1))} e^{\theta n \mathcal{I}(x_2;y_2|g(z_1))} p_{{z}}(z) \mathrm{d}z_2 \mathrm{d}z_1 \bigg ),
\end{align*}
and 
\begin{align*}
&C_2(\theta,z_2^{\star}) \notag \\
& = \frac{-1}{\theta n} \log_e \bigg (\int_0^{\infty} \int_{0}^{z_2^{\star}} e^{-\theta n \mathcal{I}(x_2;y_2|g(z_1))} p_{{z}}(z) \mathrm{d}z_2 \mathrm{d}z_1 \notag \\
& +  \int_0^{\infty} \int_{z_2^{\star}}^{\infty} e^{- \theta n \mathcal{I}(x;y|z_1,g(z_1))} e^{\theta n \mathcal{I}(x_1;y_1|z_1)} p_{{z}}(z) \mathrm{d}z_2 \mathrm{d}z_1 \bigg ),
\end{align*}
where $n = TB$. Let $\mathcal{B}(\hat{z_2}) = \lambda_1 C_1(\theta,\hat{z_2}) + \lambda_2 C_2(\theta,\hat{z_2})$, where $\hat{z_2} = z_2^{\star} + e\xi$, $z_2^{\star}$ is the optimal decoding function that solve the optimization problem (\ref{obtimization_objective}), $e$ is a constant and $\xi$ represents an arbitrary deviation. Consequently, the following condition should be satisfied\cite{arfken2013mathematical}:
\begin{equation}
\label{condition}
\frac{d}{de}\mathcal{B}(\hat{z_2}) \bigg |_{e = 0} = 0.
\end{equation}
By noting that this condition holds for any $\xi$ and that $\frac{d \hat{z_2}}{d e} = \xi$, solving (\ref{condition}) results in the following: 

\begin{align}
\label{eq_1}
& e^{- \theta n \mathcal{I}(x;y|z_1,z_2^{\star})} \bigg \{ \frac{-\lambda_1}{\psi_1} e^{\theta n \mathcal{I}(x_2;y_2|z_2^{\star})} + \frac{\lambda_2}{\psi_2} e^{\theta n \mathcal{I}(x_1;y_1|z_1)} \bigg\} \notag \\
& = \frac{\lambda_2}{\psi_2} e^{-\theta n \mathcal{I}(x_2;y_2|z_2^{\star})} - \frac{\lambda_1}{\psi_1} e^{- \theta n \mathcal{I}(x_1;y_1|z_1)}.
\end{align}
Now, let us denote $\mathcal{I}_{12} = \mathcal{I}(x;y|z_1,z_2^{\star})$, $\mathcal{I}_1 = \mathcal{I}(x_1;y_1|z_1)$ and $\mathcal{I}_2 = \mathcal{I}(x_2;y_2|z_2^{\star})$. Consequently, we can express (\ref{eq_1}) as

\begingroup
\allowdisplaybreaks
\begin{equation}
\label{eq_2}
\begin{aligned}
& e^{- \theta n \mathcal{I}_{12}} \bigg \{ -\psi_2 \lambda_1 e^{\theta n \mathcal{I}_2} + \psi_1 \lambda_2  e^{\theta n \mathcal{I}_1} \bigg\} = \\
& \hspace{3.4cm} -\psi_2 \lambda_1 e^{-\theta n \mathcal{I}_1} + \psi_1 \lambda_2 e^{- \theta n \mathcal{I}_2}.
\end{aligned}
\end{equation}
\endgroup
Let us further define $A = e^{-\theta n \mathcal{I}_2}$ and $D = e^{-\theta n \mathcal{I}_1}$. Then, (\ref{eq_2}) can be rewritten as 
\begin{equation*}
e^{- \theta n \mathcal{I}_{12}} \bigg \{ \frac{-\psi_2 \lambda_1}{A} + \frac{\psi_1 \lambda_2}{D}\bigg \} = - \psi_2 \lambda_1 D + \psi_1 \lambda_2 A, 
\end{equation*}
which can be further simplified as 
\begin{equation}
\label{eq_3}
e^{- \theta n \mathcal{I}_{12}} = A D = e^{-\theta n \{\mathcal{I}_1 + \mathcal{I}_2\}}.
\end{equation}
Note that (\ref{eq_3}) implies that $ \mathcal{I}_{12} = \mathcal{I}_1 + \mathcal{I}_2$ which is equivalent to having 
\begin{equation*}
\label{eq_4}
\mathcal{I}(x;y|z_1,z_2^{\star}) = \mathcal{I}(x_1;y_1|z_1) + \mathcal{I}(x_2;y_2|z_2^{\star}).
\end{equation*}

\end{thisnote}  
\bibliographystyle{IEEEtran}
\bibliography{IEEEabrv,References}

\begin{thebibliography}{10}
\providecommand{\url}[1]{#1}
\csname url@samestyle\endcsname
\providecommand{\newblock}{\relax}
\providecommand{\bibinfo}[2]{#2}
\providecommand{\BIBentrySTDinterwordspacing}{\spaceskip=0pt\relax}
\providecommand{\BIBentryALTinterwordstretchfactor}{4}
\providecommand{\BIBentryALTinterwordspacing}{\spaceskip=\fontdimen2\font plus
\BIBentryALTinterwordstretchfactor\fontdimen3\font minus
  \fontdimen4\font\relax}
\providecommand{\BIBforeignlanguage}[2]{{%
\expandafter\ifx\csname l@#1\endcsname\relax
\typeout{** WARNING: IEEEtran.bst: No hyphenation pattern has been}%
\typeout{** loaded for the language `#1'. Using the pattern for}%
\typeout{** the default language instead.}%
\else
\language=\csname l@#1\endcsname
\fi
#2}}
\providecommand{\BIBdecl}{\relax}
\BIBdecl

\bibitem{tao2012overview}
X.~Tao, X.~Xu, and Q.~Cui, ``An overview of cooperative communications,''
  \emph{{IEEE} Commun. Mag.}, vol.~50, no.~6, pp. 65--71, 2012.

\bibitem{biglieri2007multiple}
E.~Biglieri and L.~Gy{\"o}rfi, \emph{Multiple Access Channels: Theory and
  Practice}.\hskip 1em plus 0.5em minus 0.4em\relax IOS press, 2007.

\bibitem{tse1998multiaccess}
D.~N.~C. Tse and S.~V. Hanly, ``Multiaccess fading channels. i. polymatroid
  structure, optimal resource allocation and throughput capacities,''
  \emph{{IEEE} Trans. Inf. Theory}, vol.~44, no.~7, pp. 2796--2815, 1998.

\bibitem{gupta2006power}
G.~A. Gupta and S.~Toumpis, ``Power allocation over parallel {G}aussian
  multiple access and broadcast channels,'' \emph{{IEEE} Trans. Inf. Theory},
  vol.~52, no.~7, pp. 3274--3282, 2006.

\bibitem{knopp1995information}
R.~Knopp and P.~A. Humblet, ``Information capacity and power control in
  single-cell multiuser communications,'' in \emph{{IEEE} Int. Commun. Conf.
  (ICC)}, vol.~1, 1995, pp. 331--335.

\bibitem{vishwanath2001optimum}
S.~Vishwanath, S.~Jafar, and A.~Goldsmith, ``Optimum power and rate allocation
  strategies for multiple access fading channels,'' in \emph{{IEEE} Veh.
  Technol. Conf. Spring (VTC-SPRING)}, vol.~4, 2001, pp. 2888--2892.

\bibitem{viswanath2001asymptotically}
P.~Viswanath, D.~N.~C. Tse, and V.~Anantharam, ``Asymptotically optimal
  water-filling in vector multiple-access channels,'' \emph{{IEEE} Trans. Inf.
  Theory}, vol.~47, no.~1, pp. 241--267, 2001.

\bibitem{harshan2011two}
J.~Harshan and B.~S. Rajan, ``On two-user {G}aussian multiple access channels
  with finite input constellations,'' \emph{{IEEE} Trans. Inf. Theory},
  vol.~57, no.~3, pp. 1299--1327, 2011.

\bibitem{lozano2006optimum}
A.~Lozano, A.~M. Tulino, and S.~Verd{\'u}, ``Optimum power allocation for
  parallel {G}aussian channels with arbitrary input distributions,''
  \emph{{IEEE} Trans. Inf. Theory}, vol.~52, no.~7, pp. 3033--3051, 2006.

\bibitem{guo2005mutual}
D.~Guo, S.~Shamai, and S.~Verd{\'u}, ``Mutual information and minimum
  mean-square error in {Gaussian} channels,'' \emph{{IEEE} Trans. Inf. Theory},
  vol.~51, no.~4, pp. 1261--1282, 2005.

\bibitem{nguyen2010outage}
K.~D. Nguyen, A.~Guillen~i Fabregas, and L.~K. Rasmussen, ``Outage exponents of
  block-fading channels with power allocation,'' \emph{{IEEE} Trans. Inf.
  Theory}, vol.~56, no.~5, pp. 2373--2381, 2010.

\bibitem{ghanem2012mac}
S.~A. Ghanem, ``{MAC} gaussian channels with arbitrary inputs: Optimal
  precoding and power allocation,'' in \emph{Int. Conf. Wireless Commun. Signal
  Process. (WCSP)}, 2012, pp. 1--6.

\bibitem{wu_negi}
D.~Wu and R.~Negi, ``Effective capacity: a wireless link model for support of
  quality of service,'' \emph{{IEEE} Trans. Wireless Commun.}, vol.~2, no.~4,
  pp. 630--643, 2003.

\bibitem{tang}
J.~Tang and X.~Zhang, ``Quality-of-service driven power and rate adaptation
  over wireless links,'' \emph{{IEEE} Trans. Wireless Commun.}, vol.~6, no.~8,
  pp. 3058--3068, August 2007.

\bibitem{gursoy}
M.~Gursoy, ``{MIMO} wireless communications under statistical queueing
  constraints,'' \emph{{IEEE} Trans. Inf. Theory}, vol.~57, no.~9, pp.
  5897--5917, Sept 2011.

\bibitem{ak_gur_4}
S.~Akin and M.~Gursoy, ``On the throughput and energy efficiency of cognitive
  {MIMO} transmissions,'' \emph{{IEEE} Trans. Veh. Technol.}, vol.~62, no.~7,
  pp. 3245--3260, Sept 2013.

\bibitem{gallager1968information}
R.~G. Gallager, \emph{Information theory and reliable communication}.\hskip 1em
  plus 0.5em minus 0.4em\relax Springer, 1968, vol.~2.

\bibitem{chang1994effective}
C.-S. Chang, P.~Heidelberger, S.~Juneja, and P.~Shahabuddin, ``Effective
  bandwidth and fast simulation of {ATM} intree networks,'' \emph{{Elsevier}
  Performance Evaluation}, vol.~20, no.~1, pp. 45--65, 1994.

\bibitem{qiao2013achievable}
D.~Qiao, M.~C. Gursoy, and S.~Velipasalar, ``Achievable throughput regions of
  fading broadcast and interference channels under {Q}o{S} constraints,''
  \emph{{IEEE} Trans. Commun.}, vol.~61, no.~9, pp. 3730--3740, 2013.

\bibitem{qiao2012transmission}
------, ``Transmission strategies in multiple-access fading channels with
  statistical {Q}o{S} constraints,'' \emph{{IEEE} Trans. Inf. Theory}, vol.~58,
  no.~3, pp. 1578--1593, 2012.

\bibitem{ozcan2014qos}
G.~Ozcan and M.~C. Gursoy, ``Qos-driven power control for fading channels with
  arbitrary input distributions,'' in \emph{{IEEE} Int. Symp. Inform. Theory
  (ISIT)}, 2014, pp. 1381--1385.

\bibitem{jindal2004duality}
N.~Jindal, S.~Vishwanath, and A.~Goldsmith, ``On the duality of {G}aussian
  multiple-access and broadcast channels,'' \emph{{IEEE} Trans. Inf. Theory},
  vol.~50, no.~5, pp. 768--783, 2004.

\bibitem{jeong2012mimo}
W.~H. Jeong, J.~S. Kim, M.-w. Jung, and K.-S. Kim, ``{MIMO} channel measurement
  and analysis for 4{G} mobile communication,'' in \emph{{Springer} Convergence
  Hybrid Inform. Technol.}, 2012, pp. 676--682.

\bibitem{tse2005fundamentals}
D.~Tse and P.~Viswanath, \emph{Fundamentals of wireless communication}.\hskip
  1em plus 0.5em minus 0.4em\relax Cambridge university press, 2005.

\bibitem{arfken2013mathematical}
G.~B. Arfken, \emph{Mathematical methods for physicists}.\hskip 1em plus 0.5em
  minus 0.4em\relax Academic press, 2013.

\end{thebibliography}

\end{document}